\documentclass[usenatbib]{mn2e}

\usepackage[english]{babel}

\usepackage{graphicx}
\usepackage{dcolumn}
\usepackage{bm}
\usepackage{epstopdf}

\newcommand{\cM}{{\cal M}}
\newcommand{\cN}{{\cal N}}

\newcommand{\bin}{{\rm bin}}
\newcommand{\nbin}{{n_\bin}}

\renewcommand{\sl}{{\it sl}}

\newcommand{\vr}{{\bf r}}

\newcommand{\myskip}[1]{}

\newcommand{\mbar}{{\overline m}}

\renewcommand{\d}{{\rm d}}

\newcommand{\BEQ}{\begin{eqnarray}}
\newcommand{\EEQ}{\end{eqnarray}}
\newcommand{\BEA}{\begin{eqnarray}}
\newcommand{\EEA}{\end{eqnarray}}
\newcommand{\nn}{\nonumber}
\newcommand{\Sigmab}{\overline\Sigma}

\newcommand{\cm}{{\rm cm}}
\newcommand{\gr}{{\rm gr}}
\newcommand{\km}{{\rm km}}

\newcommand{\s}{{\rm s}}

\newcommand{\kpc}{{\rm kpc}}

\newcommand{\LCDM }{$\Lambda$CDM}

\begin{document}

\title{Modified Gravity (MOG) and its test on galaxy clusters}

\author[Nieuwenhuizen, Morandi and Limousin]
{Theodorus M. Nieuwenhuizen$^{1,2}$\thanks{E-mail: t.m.nieuwenhuizen@uva.nl},
 Andrea Morandi$^{3}$\thanks{E-mail: andrea@wise.tau.ac.il} ,
 Marceau Limousin $^{4}\thanks{E-mail: aap}$ \\
\\ $^{1}$Institute for Theoretical Physics, University of Amsterdam, Science Park 904, P.O. Box 94485,  
1090 GL  Amsterdam, The Netherlands \\
$^{2}$ International Institute of Physics, UFRG, Anel Vi\'ario da UFRN - Lagoa Nova, Natal - RN, 59064-741, Brazil \\
 $^{3}$ Physics Department, University of Alabama in Huntsville, Huntsville, AL 35899, USA \\
 $^4$  Aix MarseilleUniversit\'e,  CNRS, LAM, Laboratoire d?Astrophysique de Marseille, Marseille, France}

\maketitle

\begin{abstract}
The MOdified Gravity (MOG) theory of J. Moffat assumes a massive vector particle which causes a repulsive contribution 
to the tensor gravitation. For the galaxy cluster A1689 new data for the X-ray gas and the strong lensing properties are presented.
Fits to MOG are possible by adjusting the galaxy density profile.
However, this appears to work as an effective dark matter component, posing a serious problem for MOG.
New gas and strong lensing data for the cluster A1835 support these conclusions and point at a tendency of the gas-alone 
to overestimate the lensing effects in MOG theory.
\end{abstract} 

\begin{keywords}
 gravitation; 
galaxies: clusters: individual:...;  gravitational lensing: strong
\end{keywords}

\section{Introduction}

With the ongoing no-show of the WIMP and the axion, and
 the natural dark matter candidate, the neutrino,  long ruled out (but not given up, see e.g.  \citet{nieuwenhuizen2016dirac}), 
the dark matter riddle is ripe for reconsideration. 
One option is that dark matter (DM) effects do not arise from some particle but from a deviation from Newton's law in the weak-gravity regime. 
Examples of modified gravity theories are:  Modified Newtonian Dynamics (MOND)
 \citep{milgrom1983modification,famaey2012modified};
Entropic Gravity (EG1)  \citep{verlinde2011origin} and Emergent Gravity (EG2) \citep{verlinde2017emergent}, which appears to be MOND-like;
so-called  $f(R)$ theories \citep{sotiriou2010f(R),defelice2010f(R)};  and MOdified Gravity (MOG) \citep{moffat2005gravitational}.
It is thus important to test these theories as much as possible.
In  \citet{nieuwenhuizen2016zwicky} one of us investigates whether these theories achieve to explain lensing properties 
of  a well documented galaxy cluster, Abell 1689 (shortly: A1689).  It stands out since it is large, heavy and probably quite relaxed. 
Good data exist for the X-ray gas and its strong and weak lensing properties \citep{Morandi2012Xray1835}.
Within the often employed spherical approximation, the investigation reveals that MOND, EG1, MOG and $f(R)$ theories  fail 
to give proper account of the lensing data; by default this also applies to EG2.
It is noted that  MOND and EG1 may survive if additional cluster DM, like $\sim 1.9$ eV neutrinos, is added.
As to the spherical approximation, let us note that the axis ratio of the gas is $ 1.1 -1.06$ (on the plane of the sky) 
and $1.5 - 1.3$ (along the line of sight),  moving from the centre toward the X-ray boundary \citep{morandi2011triaxiality}. 
Triaxial studies of this cluster have more recently been conducted by \citet{umetsu2015three}.

The reported failure of MOG invoked a reaction by Moffat and Zoolideh Haghighi (MZH)  who conclude that acceleration data 
of the A1689 cluster fare well within MOG \citep{moffat2016modified}.  
 \citet{hodson2017galaxy}, on the other hand, seek to change MOND to incorporate an extra effect in clusters,
 and also compare to EG and MOG.
 
Because of the high stakes of the issue, we return here to the situation within the spherical approximation.
Our reaction involves several points. First of all, it goes without saying that if the MOG acceleration predictions indeed fit the measurements while 
lensing data fail to do so, MOG remains a problematic theory.
Second,  to the best of our knowledge, there does not exist explicit acceleration data for A1689. The data points of figure 2 of \citet{nieuwenhuizen2016zwicky}
are estimates, and partly upper estimates,  for the acceleration in theories, such as MOG, where light moves 
in the gravitational potential  \citep{nieuwenhuizen2016zwicky}.
The vanishing of the MZH acceleration at small radii in their figure 1 is perfectly physical while consistent with the
 finite value of their upper bound.
Third,  MZH employ A1689 parameters from our paper \citet{nieuwenhuizen2016zwicky}, in particular from 2 runs of X-ray data  by the Chandra satellite
that were introduced by us in   \citet{nieuwenhuizen2011prediction}.
Below we present here the final A1689 Chandra data for the X-ray gas, and notice a calibration error in analysing the previous data sets. 
Hence the gas fits must be redone; the implication for MOG will be presented below.

We also present new gas and strong lensing data for a second, well relaxed cluster,  A1835,  and analyse them in a similar fashion.

In section 2 we describe the final CHANDRA data for A1689 and new gas data for A1835, and fit them to analytical formulas.
We also present new strong lensing data for both clusters.
In section 3 we recall some relations between observables.
In section 4 we present MOG theory and in section 5 the comparison with  the A1689 and A1835 data.
We close with a discussion.

\section{Data description}

\subsection{Abell 1689}

All our results are scaled to the flat \LCDM \  cosmology with  $\Omega_M= 0.3$,    $\Omega_\Lambda= 0.7$, 
and a  Hubble constant $H_0 = 70\,h_{70}$ km/s Mpc with $h_{70}=1$.
At the redshift $z= 0.183$ of the A1689 cluster,  $1''$ corresponds to 3.076 kpc.

\subsubsection{Data for the X-ray gas}

We present the final data of the  CHANDRA X-ray Observatory. 
The data reduction was carried out using the CIAO 4.8.1 and Heasoft 6.19 software suites, 
in conjunction with the Chandra calibration database (CALDB) version 4.7.2.
Figure 1 exhibits the resulting 56 data CHANDRA points at radii between 7.7 and 963 kpc.

Let us recall some properties of the X-ray gas.
We adopt a typical $Z=0.3$ solar metallicity for A1689, so that to a good approximation $n_p =11\,n_\alpha$. 
With elements heavier than He neglected and 25\% of the gas weight in He, this implies that $n_e=(13/11)n_p$.
The particle density is $n_e+n_p +n_\alpha=(25/13)n_e$ and hence the thermal pressure $p_g =(25/13)n_ek_BT_g$. 
The mass density reads $\rho_g =m_Nn_p+4m_N n_\alpha= (15/11)m_Nn_p$, so $\rho_g=\mbar_Nn_e$ with 
$\mbar_N=(15/13)m_N=1.154\, m_N$.  The mean molecular weight in $p=\rho_gk_BT/\mu m_N$ is $\mu=3/5$.
These factors agree within a per mille with $n_e+n_{ion}=1.9254\,n_e$ and $\mu=0.5996$ 
from the solar abundance tables of  \citet{asplund2009chemical}.

Further data for $n_e$ have been obtained from the ROSAT satellite with its
Position Sensitive Proportional Counters (PSPC) camera \cite{eckert2012gas}. 
A resulting set of 50 ``parametric'' data points for $n_p$  is publicly available \citep{eckert2012gas,eckert-site}.
As  $r_i$-values we take the mids of their bins.
As seen in figures 1 and 2, the Chandra and PR data sets overlap
within their error bars for radii between 268 and 872 kpc  (except for the outlying last Chandra point). 

\subsubsection{Fit to the X-ray gas data in A1689}

 The S\'ersic mass profile $\rho=\rho_0\exp[-(r/R_g)^{1/n_g}]$ gave inspiration for a cored S\'ersic electron density 
 profile \citep{nieuwenhuizen2016dirac}, 
\BEQ \label{neSfit}
n_S(r)&=&n_e^0\exp\Big[k_g-k_g\Big(1+\frac{r^2}{R_g^{2}}\Big)^{1/(2n_g)}\Big] .
\EEQ 
The best fit of this profile to the final Chandra data gives
$\chi^2/\nu=0.692$ for all 56 points included (so that the number of degrees of freedom is $\nu=52$) 
and $\chi^2/\nu=0.504$ when the outlying last point is discarded, which we do from now on. 
The best fit for the latter case is (as expected, both cases coincide within the error bars)
\BEQ \label{nefitnotial}
n_e^0&=0.0431 \pm 0.0017\,\cm^{-3}, \quad  & k_g= 2.50 \pm 0.36 ,   \nn \\
R_g&=24.6 \pm 2.9 \, \kpc, \qquad\quad  \quad & n_g=3.32 \pm 0.21 . 
\EEQ
The statistics of the final Chandra data is better than for the two runs we employed before; even though the error bars are smaller,
the  $\chi^2$ value becomes noticeably smaller. Nevertheless, the fit (\ref{nefitnotial}) does not get essentially smaller error bars; 
we attribute this to non-sphericalities  in the cluster.
But do notice that the value of $n_e^0$ in (\ref{nefitnotial}) is a factor 1.5 smaller than the value employed in our earlier works 
due to a calibration error there.
This new value for $n_e^0$ will shift our previous fits, but appear to have no qualitative impact for MOG or other theories.

The cored S\'ersic profile has a stretched exponential decay, matched by the inner data of ROSAT/PSPC, 
but the latter data extend beyond 1 Mpc, where they expose a slower decay.
To model this, we consider two forms of a tail. First, the cored isothermal tail for the electron and mass densities,
\BEQ \label{NeTailIso}
n_T=\frac{d_tn_e^0}{r^2 + R_t^2},\quad \rho_{g,T}=\mbar_N n_T=\frac{ \sigma_g^2}{2\pi G(r^2+R_t^2)} ,
\EEQ
with $\sigma_g^2=2\pi G\mbar_Nd_tn_e^0$, is combined with the S\'ersic profile as
 \BEQ \label{nefit}
 n_e(r)=\Big(\frac{n_S^{s_t}(r)+n_T^{s_t}(r)}{1+(d_t/R_t^2)^{s_t}} \Big)^{1/s_t} .
 \EEQ
 so that $n_e(0)=n_e^0$. 
The fit leads to the very small $\chi^2(n_e)/\nu=0.32$ and the parameters
\BEQ\label{gas-pars-Iso}
&& n_e^0=0.04376 \pm 0.00098\,\cm^{-3}, \quad
k_g=2.06 \pm 0.12 , \nn \\
&& R_g=21.8 \pm 1.4 \, \kpc,\qquad \qquad \quad
n_g=3.044 \pm 0.062 ,  \nn \\
&& d_t=6660 \pm 255 \, \kpc^2, \qquad \qquad \,\,\,
\sigma_g=476.5\pm 7.7\, \km/\s,
\nn\\ &&  R_t=718 \pm 108 \,\kpc , \hspace{1.8cm} s_t=8.4 \pm 2.7.  
 \EEQ 
The relative errors in these parameters are  0.022,  0.057, 0.062,  0.021, 0.038, 0.016, 0.15, and 0.32, respectively.
Not all tail parameters are strongly constrained:  $R_t$ and $s_t$ have appreciable errors. 
One reason for this is that $s_t$ is only determined by the few data in the cross over region from $n_S$ to $n_T$. 
 The fit  for $n_e$ is exposed in figures 1 and 2. The ratios $n_{e,i}/n_e(r_i)$ are exposed in figure 2;
they overlap nearly all within their error bars with the ideal value 1, causing the small $\chi^2(n_e)/\nu=0.32$.

 \begin{figure}
\label{nedata1689}
 \includegraphics[scale=0.9]{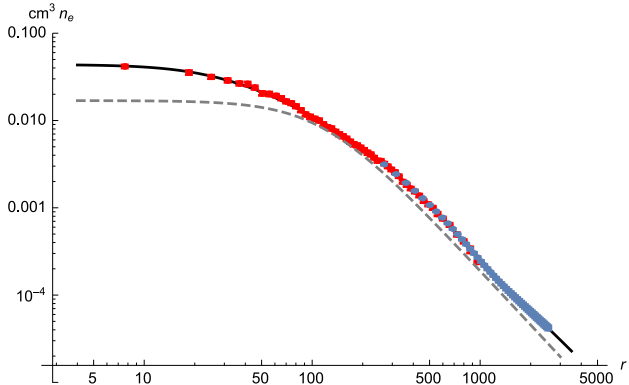}
\caption{Data for the electron density $n_e$ in A1689 from Chandra (red) and Rosat (blue), with the analytic fit 
of Eqs. (\ref{neSfit}), (\ref{NeTailIso}), (\ref{nefit})  and (\ref{gas-pars-Iso}) (full line).
Dashed line: the $\beta$-model of \citet{brownstein2006galaxy}. }
\end{figure}

\begin{figure}
\label{nereldata1689}
 \includegraphics[scale=0.9]{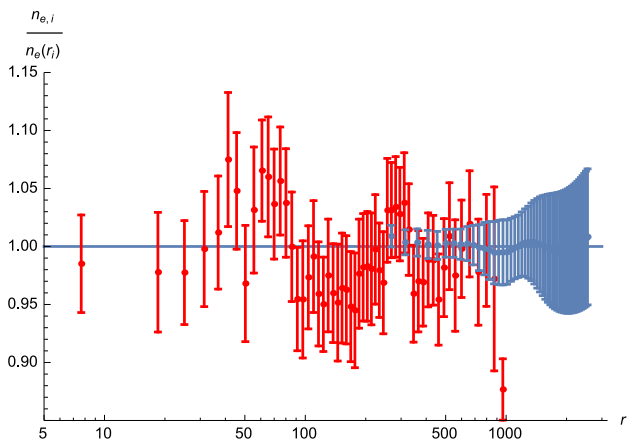}
\caption{The data for $n_e$ of figure 1 relative to the fit.
Except for the outlying last Chandra point, nearly all points lie within one standard deviation from the fit.}
\end{figure}

As  a second model we consider the Burkert tail
\BEQ \label{Burkert-tail}
n_T=\frac{d_tn_e^0R_t}{ (r+R_t)(r^2 + R_t^2)} .
\EEQ
Compared to the tail (\ref{NeTailIso}) it decays quicker and leads to an only logarithmically divergent gas mass. 
It respects the data points equally well, since it also achieves $\chi^2/\nu=0.32$. The fit parameters are
\BEQ
&& n_e^0=0.0438\pm 0.0010 \,\cm^{-3}, \quad k_g=2.04 \pm 0.12 ,  \nn \\
&& R_g=21.8 \pm 1.4 ,\cm^{-3}, \qquad \quad \, n_g=3.031 \pm 0.065,  \nn \\
&& d_t=22052 \pm 594\,\kpc^2, \qquad \quad  R_t=1750 \pm 135\,\kpc,\,\,\,   \nn\\
&& s_t=4.9 \pm 1.1.  
\EEQ
The relative errors are  here 0.023, 0.060, 0.063, 0.021, 0.027, 0.077, and  0.22,  respectively.

The $\beta$-model $n_e(r)=n_e^0(1+r^2/R_g^2)^{-3\beta/2}$ yields a considerably worse best-fit, $\chi^2/\nbin=6.8$ 
with $n_e^0=0.0243 \pm 0.0010$, $R_g=105.4 \pm 3.4\,\kpc$ and  $\beta=0.6701 \pm 0.0054$,
having the small relative errors  0.043, 0.032, and 0.0080, despite the large $\chi^2$.
The $\beta$-model employed in \citet{brownstein2006galaxy}  has parameters $\rho_0 = 0.33\, 10^{-25} \gr \, \cm^{-3}$,
corresponding to $n_e^0=0.0169\, \cm^{-3}$,   $R_g= 114.8\, \kpc$ and $\beta = 0.690$. This leads to a truly bad fit
indicated by $\chi^2/(55+50)= 85.1$, so this model can only be used with proper care.
It is depicted by the dashed line in figure 1.

\subsubsection{Strong lensing in A1689}

The strong lensing (SL) data arise from background galaxies lensed by the cluster.
While a full Einstein ring does not occur, galaxies not-too-far from the sightline to the cluster centre are observed as 
an arclet or a set of $n\le 7$ arclets in A1689. Its SL mass model was first derived in \cite{limousin2007combining}.
From the arclets the computer code Lenstool, presented by \cite{jullo2007bayesian} and \cite{kneib2011lenstool}, 
has now produced candidate maps  for the $2-d$ mass distribution;
this being an underdetermined problem, a set of maps, labeled by $\mu=1,\cdots , \cN$, can be generated.
In total $\cN=1001$ solutions (``samples'') have been achieved.
These maps are integrated over circles around the centre to yield the $2d$ mass $\cM_{2d}^{(\mu)}(r_n)$ within a cylinder of radius $r_n$.
A number of $N= 149$ radii $r_n$ are chosen such that the $\log r_n$ have uniform spacing  $0.0380$ between 
$r_1=3.15355$ kpc and $r_{149}=876.783$ kpc.
From each $\cM_{2d}^{(\mu)}$  one gets $\Sigmab_n^{(\mu)}=\cM_{2d}^{(\mu)}(r_n)/\pi r_n^2$.
Summation over $\mu$ brings the statistical averages $M_{2d}(r)=({1}/{\cN})\sum_{\mu} \cM_{2d}^{(\mu)}(r_n)$ and

\BEQ \Sigmab_n=\frac{M_{2d}(r_n)}{\pi r_n^2}=\frac{1}{\cN}\sum_{\mu} \Sigmab_n^{(\mu)}, \qquad (n=1,\cdots,149),\EEQ
In shells without arclets, the Lenstool program produces constant values for $M_{2d}$, of which only the one at smallest $r$ 
provides physical information. Hence not all $r_n$ contain proper information about $\Sigmab_n$, but
in total 117 of them do so, see figure 4.

We shall not consider related data for the line-of-sight mass density $\Sigma$, since they contain no new information.
Moreover, to obtain them from $M_{2d}(r)=\pi r^2\Sigmab(r)=2\pi\int_0^r{\rm d} s\,s\Sigma(s)$, a numerical differentiation 
is needed, which introduces ambiguities.

The  correlations due to sample-to-sample variations are
\BEQ
\Gamma_{mn}=\frac{1}{\cN}\sum_{\mu} [\Sigmab_m^{(\mu)} -\Sigmab_m] [\Sigmab_n^{(\mu)}-\Sigmab_n] .
\EEQ
The standard estimate for the error bar in $\Sigmab_n$ is $\delta\Sigmab_n=(\Gamma_{nn})^{1/2}$; 
as usual in cases of correlated data, the full information on errors is coded in the covariance matrix $\Gamma_{mn}$.

\subsection{Abell 1835}
The cluster Abell 1835 is a massive  cluster which shows several indications of a well-relaxed dynamical state.
At its redshift $z= 0.253 $,   3.947 kpc corresponds to $1''$.

\subsubsection{The X-ray gas}

The setup for A1835 presented by two of us \citep{morandi2012x} is followed.
The data reduction is carried out using the CIAO 4.8.1 and Heasoft 6.19 software suites, in conjunction with the Chandra calibration database 
(CALDB) version 4.7.2. We  measure the emission measure profile EM $\propto \int n_e^2 \d l$ from the X-ray images. 
The radial EM profile is derived with the vignetting correction and direct subtraction of the 
Cosmic X-ray Background (CXB)+particle+readout artifact background. 
For the particle background modeling, we use the scaled stowed background. 
In order to measure the CXB, we used the regions free of the source emission. 
We then deprojected the measured temperature and EM profiles in order to infer the gas density profiles.

A set of 40 data points $(r_i,n_{e,i},\delta n_{e,i})$ for the electron density has been produced.
The errors $\delta n_{e,i}$ are larger than in the A1689 case, and constrain the fit profiles less well.
We find that the following profile explains the data well,

\BEQ \label{nefit1835}
n_e(r)=n_e^0 \frac{1+r^2/R_0^2}{ (1 + r^2/R_1^2) (1 + r^2/R_2^2) }  .
\EEQ
The best fit has parameters

\BEQ \label{nefitpars1835}
n_e^0&=& 0.0927\pm 0.0070 \, \cm^{-3},\quad
R_0 = 91 \pm 13 \, \kpc, \nn\\
R_1 &=& 31.8\pm  2.9   \, \kpc,\quad
R_2 = 169 \pm 15  \, \kpc.
\EEQ
It has  $\nu=40-4$ degrees of freedom and $\chi^2/\nu=0.0778$. This stunningly low value reflects that the fit goes through nearly
all data points,  in the presence of the somewhat large error bars, see figure 3.

With $\sigma_g^2=2\pi G\mbar_Nn_e^0R_1^2R_2^2/R_0^2$, the gas mass density may be written as

\BEQ
\rho_g=\mbar_N n_e=
\frac{\sigma_g^2}{2\pi G}\,
\frac{r^2+R_0^2}{(r^2+R_1^2)(r^2+R_2^2)}.
\EEQ
It has an isothermal decay $\rho_g\approx \sigma_g^2/2\pi G r^2$ with 
\BEQ
\sigma_g=496.9\pm 6.4\,\km/\s.
\EEQ

\begin{figure}
\label{nedata}
 \includegraphics[scale=0.9]{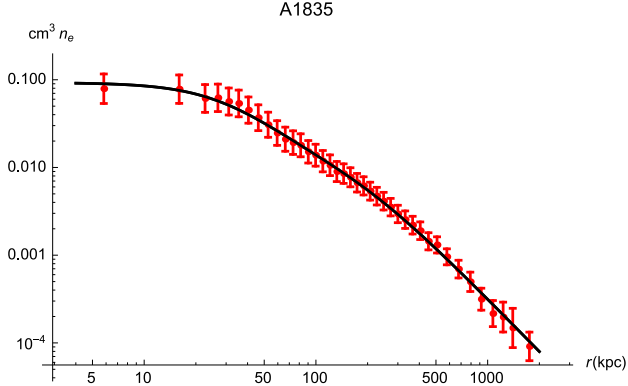}
\caption{Data for $n_e$ in A1835 and their best fit (\ref{nefit1835}) with (\ref{nefitpars1835}). }
\end{figure}

\subsubsection{Strong lensing by A1835}

The SL data have been generated in the same way as for A1689. 
The selected radii have the same spacing 0.0380 on a logarithmic scale.
The first radius is $r_1=4.027$ kpc and the last one is $r_{149}=1120$ kpc.
With again $\cN=1001$ samples, the averages $\Sigmab_n$ have been determined
in the way described in section 2.1.3. Also here 117 of the $r_n$ contain information.

\section{Lensing observables}

Because the background galaxies are far removed from the cluster, the lensing effects can be thought of as occurring due to mass
projected onto the plane through the cluster centre. 
One studies the $2d$ mass $\cM_{2d}$, that is, the mass contained in a cylinder of radius $r$ around the sight line.
Its average over the disk is  $\Sigmab(r)=\cM_{2d}(r)/\pi r^2$. 
This quantity can be expressed in terms of the $3d$ mass density,
\BEQ \label{Sb1int2}
\Sigmab(r)=\frac{4}{r^2}\int_0^r\d s\,s^2\rho(s)+\int_r^\infty\d s\,\frac{4s\rho(s)}{s+\sqrt{s^2-r^2}}.
\EEQ
In modified gravity $\cM_{2d}$ is an effective mass and $\rho$ an effective mass density. 
In general $\Sigmab$ can be expressed in terms of the gravitational potential $\varphi$ as \citep{nieuwenhuizen2009non}
\BEQ\label{Sigmabphip}
 \Sigmab(r)=
\frac{1}{\pi G} \int_0^\infty{\rm d} s\, \varphi'(r\cosh s) .
\EEQ
This expression  holds not only for general relativity but for any theory in which light moves in the gravitational potential or, at least,
does so in the first post-Newtonian approximation. In particular, it applies to MOG.

\section{MOdified Gravity (MOG)}

The MOdified Gravity theory of J. Moffat aims to replace dark matter by a modification of Newton's law  \citep{moffat2006scalar}.
Next to the standard tensor field, there is a massive vector field, which adds a repulsive term to the gravitational potential.
In MOG the potential reads
\BEQ
\varphi
&=&-G\int\d^3r' \frac{\rho_m(\vr ')}{|\vr-\vr'|}\Big(\alpha +1-\alpha e^{-\mu|\vr-\vr'|} \Big) \nn \\
&=&(\alpha+1)\varphi_N+\alpha\varphi_V  
\label{phiNV=}
\EEQ
with the subscript $N$ denoting ``Newton'' and $V$ ``vector''.
$\alpha$ is a dimensionless parameter relating the strength of the tensor field to Newton's constant $G$ and $\mu$ is the inverse range of the vector field.
A fit to galaxy catalogs yields $\alpha=8.89\pm0.34$ and $\mu=0.042\pm0.004$ kpc$^{-1}$  \citep{moffat2013mog}.
These errors are small enough to have no influence on our conclusions determined by the central values.

The separate parts of the potential satisfy the massless and massive Poisson equations, respectively,
\BEQ \label{dv1dv2}
&&
\nabla^2 \varphi_N =4\pi G\rho_m , \\
&& \label{dv2dv1}
\nabla^2\varphi_V-\mu^2\varphi_V=-4\pi G\rho_m ,
\EEQ
In the philosophy that only baryonic matter exists, the matter density consists of galaxies and X-ray gas, 
\BEQ
\rho_m=\rho_G+\rho_g.
\EEQ

In case of spherical symmetry we may introduce
\BEQ
J(r)=4\pi G \,r\rho_m(r), 
\EEQ
and derive the explicit expressions
\BEQ \label{phiMOG}
\varphi(r)&=&-\int_0^r\d u\,\Big[(\alpha+1)u-\frac{\alpha}{\mu} e^{-\mu r}\sinh \mu u \Big]\frac{J(u)}{r}
\nn\\ && \! -\int_r^\infty\d u\,\Big[(\alpha+1)r-\frac{\alpha}{\mu}  e^{-\mu u}\sinh \mu r \Big]\frac{J(u)}{r}.
\EEQ
Here the terms proportional to $\alpha+1$ are Newtonian, while the ones proportional to $\alpha$
are derived by writing Eq. (\ref{dv2dv1}) as $(r\varphi_V)''-\mu^2\,r\varphi_V=-J(r)$ and employing
 the Greens function $\exp(-\mu r_>)(\sinh \mu r_<)/\mu $, where $r_<=$ min$(r,u)$ and  $r_>=$ max$(r,u)$.
As it should,  employing the decomposition (\ref{phiNV=}), 
Eq. (\ref{phiMOG}) may be checked from Eqs. (\ref{dv1dv2}), (\ref{dv2dv1}).

For large $r$ it follows that \citep{moffat2006scalar}

\BEQ
\label{phi-approx}
\varphi(r) &\approx&(\alpha+1)\phi_N(r)+\frac{4\pi\alpha G }{\mu^2}\rho_m(r) \nn\\
&\approx&(\alpha+1)\phi_N(r).
\EEQ
The acceleration is inwards and has magnitude
\BEQ
\varphi'(r)&=&\frac{\alpha+1}{r^2}\int_0^r\d u\,u J(u)  \nn\\
&-&\frac{\alpha}{\mu r^2} (1+\mu r)e^{-\mu r}\int_0^r\d u\,\sinh\mu u \, J(u) \\
 &+&\frac{\alpha}{\mu r^2 } (\mu r\cosh  \mu r-\sinh\mu r )\int_r^\infty \d u\,e^{-\mu u}\, J(u) .\nn
\EEQ
Its small-$r$ behaviour reads $\varphi'(r) =Cr$ with
\BEQ\hspace{-0.5mm}
\label{MOGsmallr}
C=\frac{4\pi G}{3}\Big[  \rho_m(0)+\alpha \mu^2 \int_0^\infty \d u\,e^{-\mu u}\, u\,\rho_m(u) \Big]\, ,
\EEQ
which is non-Newtonian since the second term is non-zero.
It will only be small if the range of $\rho$ is much smaller than $1/\mu$, like for stars and their planetary systems.

\section{MOG applied to Clusters}

\subsection{A large set of clusters}

A set of 11 clusters is analysed in \citet{moffat2014mog} and a set of 106 clusters in \citet{brownstein2006galaxy}. 
Gas is modelled by $\beta$-profiles (not necessarily an optimal fit, see figure 1)
while hydrostatic equilibrium is assumed (now known to be often violated in the outskirts).
Most of these clusters are non-relaxed, non-spherical and not well documented, e.g., lacking data for the X-ray gas. 
With the resulting model parameters not well constrained, these fits can at best be indicative.
Conclusive indications must necessarily derive from well constrained cases.

\subsection{Abell 1689}

\newcommand{\ul}{\underline}

We first consider the application to A1689. 
The new $\Sigmab$ data with their error bars taken from the diagonal elements of the covariance matrix, are presented in figure 4.
The MOG contribution of the new X-ray gas data alone, that is, of the gas in the absence of galaxies, is depicted 
by the dashed lines in figure 4, corresponding to the isothermal and Burkert tails, respectively. 
It is seen that  MOG predicts approximately the proper strength for $\Sigmab$ at $r\sim1$ Mpc,
but deviates quickly at lower $r$. To achieve a matching of the data, 
a tentative fit for the galaxy distribution, inspired by the one of \citet{limousin2007combining}, is provided by
\BEQ
\rho_G=\frac{6 (1 + r^2/R_0^2)}{ (1 + r^2/R_1^2)(1 + r^2/R_2^2) (1 + r^2/R_3^2)^{0.4} }\,\frac{m_N}{\cm^3},
\EEQ
with $\{R_0,R_1,R_2,R_3\}=\{10,4,15,130\}\,\kpc$.  Its effect on $\Sigmab$ is presented in figure 4.
Its slow decay factor $(1 + r^2/R_3^2)^{-0.4}$ expresses that this $\rho_G$
 tries to use the galaxy  distribution as an effective a dark matter component,
a behavior that goes against the philosophy of MOG and was encountered previously \citep{nieuwenhuizen2016zwicky}.
The mass in galaxies  is $5.3\times10^{12}M_\odot$ within 100 kpc and further $2.2\times10^{13}M_\odot$
between 100 kpc and 1 Mpc, while the gas mass is  $8.0\times10^{13}M_\odot$ in the latter domain.
Fits with such a fraction of baryonic mass in galaxies are not acceptable, often the brightest cluster galaxy
is considered to dominate the combined mass of the galaxies.

\begin{figure}
\label{SiBMOG1689}
 \includegraphics[scale=0.9]{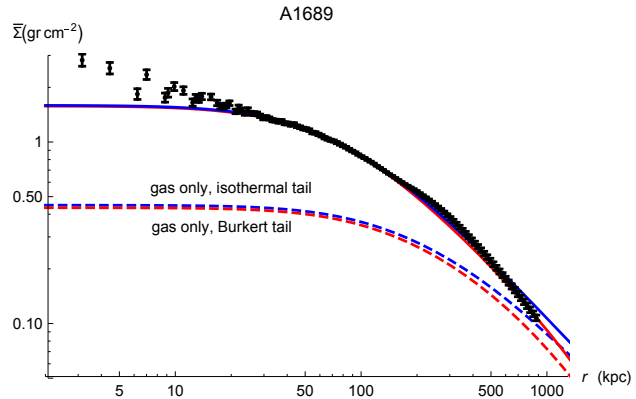}
 \caption{$\Sigmab(r)$ in A1689 in MOG theory. The dashed lines show the contribution of the X-ray gas, with isothermal 
 tail (blue) or Burkert tail (red).  The full lines exhibit the addition of the galaxy mass density in both cases.
They mimic a dark matter component.}
 \end{figure}

 \subsection{Abell 1835}
 
The $\Sigmab$ data are presented in figure 5, together with the effect of the gas alone.
A tentative match with the data is found for the galaxy mass density profile

\BEQ
\rho_G=\frac{5.3\, (1 + r^2/R_0^2)^2}{(1 + r^2/R_1^2)^2(1 + r^2/R_2^2)^2} \frac{m_N}{\cm^3} ,
\EEQ
with 

\BEQ
R_0=88\,\kpc,\quad 
R_1=11\,\kpc,\quad 
R_2=570\,\kpc.
\EEQ
The result is also presented in figure 5.
Because $R_2$ is large, this profile again acts as a form of dark matter.
The galaxies' mass is $2.0\times10^{12}M_\odot$ within 100 kpc;
there should still be $3.0\times10^{13} M_\odot$ in galaxies between 100 kpc and 1 Mpc, 
to be compared with $9.2\times10^{13}M_\odot$ in gas in that domain. Such a large fraction of bayrons in galaxies is unrealistic.

In the large $r$ domain ($r>700$ kpc) the gas already produces a larger $\Sigmab$ than deduced from the lensing alone;
this impossibility is indeed worrisome because of the different trends, so that intersection between data and the gas contribution must occur. 
Figure  5 exhibits this behavior for the best gas fit, with an 1--$\sigma$ error band in the amplitude.
A similar but less pronounced behavior is present for A1689 with an isothermal tail to the gas data, see figure  4.
The present data thus point at a serious problem for MOG.
From the anti-MOG, pro-dark-matter perspective, it may simply express that MOG's large-$r$ enhancement factor  
with respect to baryons, $\alpha+1=9.9$, should not replace the standard cosmic total-to-baryonic matter density ratio 
$\Omega_c/\Omega_B+1\approx 6.4$.
This then implies that a large portion of the mass is still missing, and this necessitates a component of
 hidden baryons or dark matter in the cluster core and outskirts.

\begin{figure}
\label{SiBMOG1835}
 \includegraphics[scale=0.9]{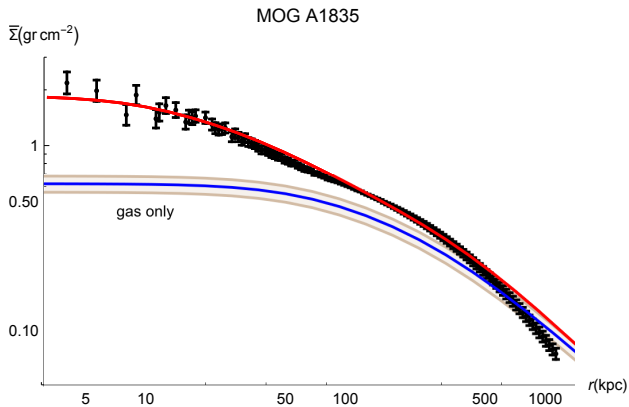}
 \caption{$\Sigmab(r)$ in A1835 in MOG theory. The lower lines show the contribution of the X-ray gas with $1-\sigma$ error bars.
 The upper line exhibits the additional effect of the galaxy mass density. It mimics a dark matter component.}
 \end{figure}

\subsection{Further aspects of MOG}

Let us speculate on other aspects of MOG.  Larger values of $\mu$ have been considered.
MZH, e.g., also consider $\mu=0.125$ kpc$^{-1}$, while \citet{DEMARTINO2017440}
investigate the scale dependence of $\mu$ and $\alpha$.
Taking a larger $\mu$ value at fixed $r$ drives MOG further away from Newton theory. 
For smooth mass distributions, 
the small-$r$ coefficient (\ref{MOGsmallr}) will converge for large $\mu$ to the non-Newtonian form

\BEQ \label{MOGsmallr2}
\varphi'(r) \to \frac{4\pi G}{3}(\alpha+1) \rho_m(0),
\EEQ
 in accord with (\ref{phi-approx}) but likely problematic in practice.

The point-mass MOG potential
\BEQ
\varphi=-\frac{GM}{r}(1+\alpha-\alpha e^{-\mu r})
\EEQ
will have the exponential vanishing in the application to satellite galaxies, and be again $\alpha+1\sim10$ times stronger than 
the Newton potential. We do not expect that smearing of the mass distributions will compensate this effect.

\section{Conclusion}

We have presented new data sets for the X-ray gas density and strong lensing effects of the well studied cluster
A1689 and the now accordingly investigated cluster A1835. These data sets are considered within MOG theory.
It is found that the gas alone matches the lensing property $\Sigmab(r)=M_{2d}(r)/\pi r^2$ around $r=1$ Mpc.
 At smaller $r$ an extra effect is needed. The demand of a mass density profile for the galaxies
 localised near the cluster centre appears to be in conflict with the demand that no dark matter is present.
 Fits tend to need matter from galaxies far from the centre, where there exist not so many of them.
 There also exists a trend for the gas to already overshoot the lensing data at large $r$, in particular for A1835, 
 which is physically impossible. On the scale of interacting satellite galaxies the gravitational potential 
seems to strongly overestimate the Newtonian value.  
 These issues pose serious problems for the MOG theory.
 
\section*{ Acknowledgements}
The authors are grateful for the careful remarks and useful hints by the unknown referee.
A. M. acknowledges support from Chandra grant GO415115X and NASA grants NNX14AI29G and NNX15AJ30G.
M. L. acknowledges CNRS, CNES and PNCG for support, 
the staff of the ``Cluster de calcul intensif HPC'' Platform of the OSU Institut Pyth\'eas 
(Aix-Marseille Universit\'e, INSU-CNRS) for providing computing facilities, and
M. Libes and C. Yohia from the Service Informatique de Pyth\'eas (SIP) for technical assistance.

 \bibliography{A1689-MOG-bib}

\bibliographystyle{mnras}

\end{document}